\documentclass[twocolumn,showpacs,preprintnumbers,amsmath,amssymb,prl]{revtex4}

\usepackage{graphicx}
\usepackage{dcolumn}
\usepackage{bm}

\begin{document}

\preprint{VERSION 323.00}

\title{Aharonov-Bohm oscillations of a tunable quantum ring}

\author{U. F. Keyser}
 \email{keyser@nano.uni-hannover.de}
 \homepage{http://www.nano.uni-hannover.de}
\author{S. Borck}
\author{R. J. Haug}
\affiliation{Institut f\"ur Festk\"orperphysik, Universit\"at Hannover, Appelstr. 2, 30167 Hannover, Germany}

\author{M. Bichler}
\author{G. Abstreiter}
\affiliation{Walter Schottky Institut, TU M\"unchen, 85748 Garching, Germany}

\author{W. Wegscheider}
\affiliation{Angewandte und Experimentelle Physik,  Universit\"at Regensburg, 93040 Regensburg, Germany}

\date{\today}

\begin{abstract}
With an atomic force microscope a ring geometry with self-aligned
in-plane gates was directly written into a
GaAs/AlGaAs-heterostructure. Transport measurements in the open
regime show only one transmitting mode and Aharonov-Bohm
oscillations with more than 50\% modulation are observed in the
conductance. The tuning via in-plane gates allows to study the
Aharonov-Bohm effect in the whole range from the open ring to the
Coulomb-blockade regime.
\end{abstract}


\maketitle

The Aharonov-Bohm (AB) effect~\cite{ab-effekt59} has attracted
much research interest over the last years in mesoscopic
semiconductor physics. Several groups realized lateral
Aharonov-Bohm rings~\cite{ismail91,pedersen00,casse00} in
heterostructures using well established techniques like electron
beam lithography and various etching techniques. Recently Piazza
et al. realized a vertical AB-interferometer which showed
oscillations with amplitudes of 30\%~\cite{piazza00}. Several
groups used AB-rings as phase detectors for the transport through
quantum dots situated in one arm of the
ring~\cite{yacoby95,vanderwiel00}. Measurements on the spectrum of
quantum rings were also performed with optical
methods~\cite{lorke00,warburton00}.

Here we present data measured on an asymmetric quantum ring with
two tuneable point contacts. The ring was fabricated using an
atomic force microscope (AFM) as nanolithographic tool. The design
of the quantum ring enables us to measure electron interference
effects and single-electron charging effects on the same device.
In the open regime (resistance of contacts $R \approx h/e^2$) the
quantum ring acts as an electron interferometer and we observe
AB-oscillations with a modulation amplitude of more than 50\%.
When the point contacts are pinched off ($R \gg h/e^2$) the ring
is separated from the leads by tunnelling barriers. Because of the
small diameter we can measure the typical Coulomb-blockade (CB)
oscillations expected for a single-electron transistor. In
addition we observe AB-like oscillations also in this regime.
\begin{figure}
\includegraphics[width=7cm]{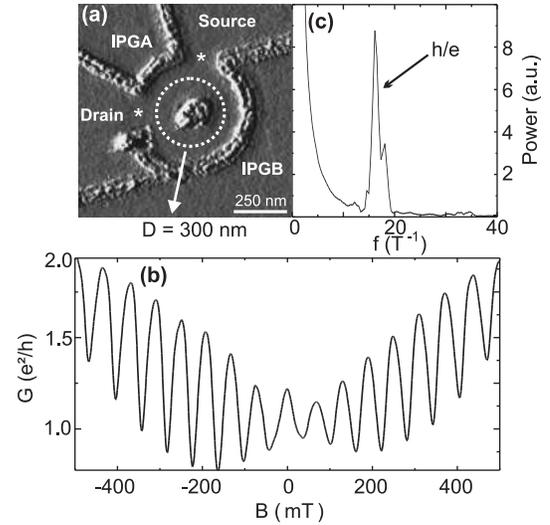}
\caption{(a) AFM-image of the ring-structure. IPGA, IPGB denote
the in-plane gates. The two point contacts are marked with an {\bf
*}. The white dashed circle (diameter 300 nm) corresponds to the
transmitted mode extracted from (b) and (c). The inner diameter of
the ring is $D_i = 190$~nm and the outer $D_o = 450$~nm. (b)
Measurements of the conductance $G$ through the ring in
perpendicular magnetic field $B$ ($V_A = 95$~mV, $V_B = 120$~mV,
$T = 25$~mK). (c) Power spectrum of the measurement in (b).}
\label{ab+fft}
\end{figure}

The fabrication was done on a GaAs/AlGaAs-heterostructure
consisting of 5~nm thick GaAs cap layer, 8~nm of AlGaAs, the
Si-$\delta$-layer, a 20~nm wide AlGaAs barrier and 100 nm of GaAs
(from top to bottom). The two-dimensional electron gas is located
34 nm below the surface with a density of 5$\cdot
10^{15}$~m$^{-2}$ and a mobility of 42~m$^2$/Vs. The mean free
path of the electrons is $4.9$~$\mu$m. A Hall-bar geometry was
defined by standard photolithography and wet-chemical etching. The
electron gas was then contacted with alloyed Au/Ge-contacts. For
the nanolithography of our samples we use local anodic oxidation
with an atomic force microscope (AFM). It was shown by Ishii and
Matsumoto that shallow 2DEGs are depleted when the surface is
oxidized using a conducting AFM-tip~\cite{ishii95}. Held et
al.~\cite{held98} demonstrated the very short depletion length
(less than 50 nm) of this process. To obtain insulating lines with
a breakdown voltage of $\pm 300$~mV we apply high oxidation
currents ($I \approx 1.0~\mu$A) and write the structures at least
twice. Thus we create a lateral structure with self-aligned
in-plane gates (IPG). For more details on our current-controlled
local oxidation see Ref.~\cite{keyser}.
\begin{figure}
\includegraphics[width=6cm]{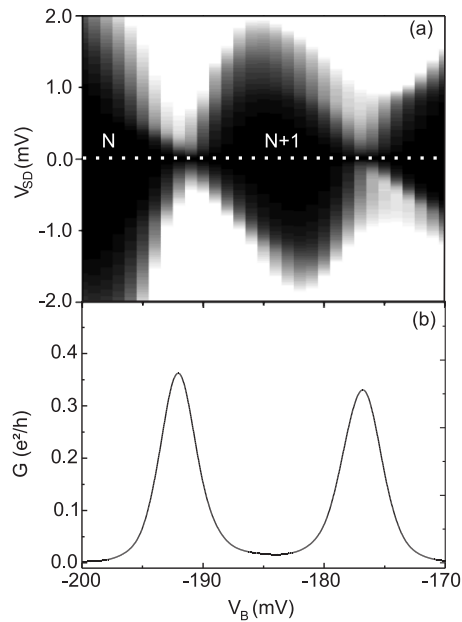}
\caption{(a) Greyscale plot of the source-drain current $I_{SD}$
as function of $V_B$ and source-drain voltage $V_{SD}$ (grid size:
$\Delta V_{SD} = 10$~$\mu$V and $\Delta V_B = 1.0$~mV). White
corresponds to $I_{SD} > 5$~nA and black to $I_{SD} < 0.1$~nA. The
Coulomb-blockade diamond indicate a charging energy of
approximately 1~meV. (b) Conductance $G$ at $V_{SD} = 0$~mV
through the quantum ring. Two Coulomb-blockade peaks are shown.}
\label{cb-osc}
\end{figure}

In Fig.~\ref{ab+fft}(a) an AFM-image of the geometry is shown. The
oxide lines appear as rough white-black surface (linewidth
$<120$~nm). With the oxide lines we define self-aligned in-plane
gates (IPG). These are labeled as IPGA, IPGB and the applied
voltages as $V_A,V_B$. We chose the inner diameter of the quantum
ring as $D_i=190$~nm and the outer $D_o=450$~nm. Thus the outer
circumference of our ring is more than three times shorter than
the mean free path in the unpatterned 2DEG. We can assume that we
are in the ballistic regime and neglect effects of elastic
scattering in the ring. The white dashed circle shows the expected
path of the electrons passing through the ring with $D \approx
300$~nm. The ring is connected to the source and drain contacts by
two 150~nm wide point contacts (marked with an *). The
conductivity through both point contacts is mainly controlled by
the voltage $V_A$ applied to IPGA. For $V_A=V_B=0$~mV the point
contacts are conducting and the resistance of the device is found
to be between $h/e^2$ and $h/2e^2$. This suggests that only one
conducting channel is contributing to the transport. All
measurements were performed in a dilution refrigerator at a base
temperature of $T = 25$~mK with a standard lock-in technique and
an AC-voltage of 5~$\mu$V ($f = 89$~Hz).

Fig.~\ref{ab+fft}(b) shows the conductance $G(B)$ of the quantum
ring in the open regime with $V_A=95$~mV and $V_B=120$~mV ($R <
h/e^2$) in a perpendicular magnetic field $B$. We observe
AB-oscillations with an amplitude of more than 50\%. The power
spectrum of the data in Fig.~\ref{ab+fft}(b) is shown in
Fig.~\ref{ab+fft}(c). In the power spectrum the dominating feature
is a frequency of 16~T$^{-1}$. This means that every 62~mT a flux
quantum enters the area enclosed by the electron paths. From this
value we can determine the diameter of the dominating electron
orbit in our ring to $D = 300$~nm. This fits perfectly for the
geometry as shown in Fig.~\ref{ab+fft}(a) by the white dashed
line. The observation of a dominant frequency at 16~T$^{-1}$ is an
additional indication that only one conducting channel is present
in this voltage regime.
\begin{figure}
\includegraphics[width=6cm]{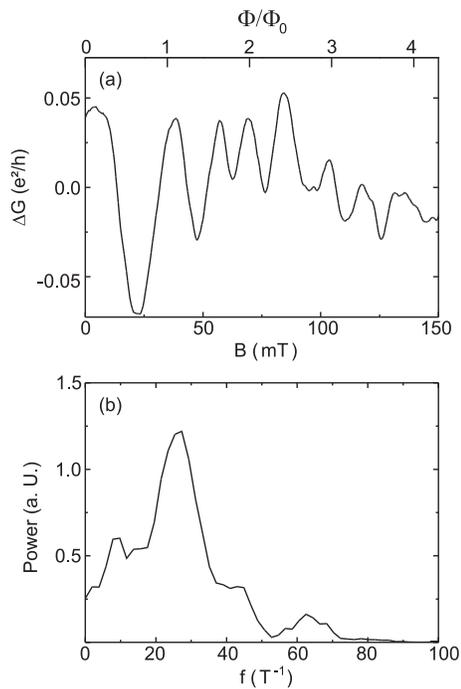}
\caption{(a) Quasi-periodic Aharonov-Bohm oscillations of the
second CB-peak in Fig.\ref{cb-osc}(b) at $V_{SD} \approx 0$~mV and
$V_A = -150$~mV and $V_B = -177$~mV. The change of peak height is
shown after substraction of an offset. (b) Power spectrum of the
data in (a). A broad distribution is observed with a maximum at
28~T$^{-1}$.} \label{cb-ab}
\end{figure}

Due to the geometry of the device it is possible to pinch off both
point contacts simultaneously with a negative voltage applied to
IPGA. With $V_A=-150$~mV we create electronic barriers at the
constrictions with low tunnelling transparency and are now able to
study the quantum ring in the Coulomb-blockade (CB) regime. In
Fig.~\ref{cb-osc}(b) two CB-peaks of the structure are shown.
$G(V_B)$ exhibits peaks for every electron added to the ring. By
sweeping IPGB between -200~mV and -170~mV we increase the electron
number $N$ on the ring to $N+2$. The two peaks are separated by
15~mV which gives a sidegate capacitance $C_B\approx 11$~aF. This
value is consistent with values of our previous
devices~\cite{keyser}.

Fig.~\ref{cb-osc}(a) shows a greyscale-plot of the source-drain
current $I_{SD}$ through the ring in the Coulomb-blockade regime.
White (black) displays high (low) current. We observe nice
Coulomb-blockade diamonds as expected for a single electron
transistor. By analyzing the width of the $N+1$-diamond in
Fig.~\ref{cb-osc}(a) we determine the overall capacitance
$C_\Sigma \approx 160$~aF corresponding to a charging energy of
the ring of approximately 1~meV.

In the CB-regime we also observe AB-oscillations. By applying a
perpendicular magnetic field we measured the change in the
conductance of the second CB-peak ($V_B = -177$~mV). The result is
shown in Fig.~\ref{cb-ab}(a) where the variation of the peak
height $\Delta G(B)$ is plotted against magnetic field $B$ after
subtraction of a slowly varying background. $\Delta G(B)$ shows a
periodic modulation with increasing magnetic field. According to
Tan and Inkson~\cite{tan96} one expects that the peak amplitude
oscillates with each flux quantum $\Phi_0$ entering the quantum
ring. This behaviour was recently observed by Fuhrer et
al.~\cite{fuhrer_cond01} for similar fabricated quantum rings. The
more complicated oscillations shown in Fig.~\ref{cb-ab}(a) can be
interpreted as a mixing of several electron orbits that contribute
to the tunnelling transport since the ground state changes with
magnetic field.

The corresponding power spectrum of $\Delta G$ in
Fig.~\ref{cb-ab}(a) is shown in Fig.~\ref{cb-ab}(b). For our ring
we obtain a clear peak centered at 28~T$^{-1}$ which means that
every 36~mT a flux quantum enters the area surrounded by the
tunnelling electrons. The corresponding electron orbit has a
diameter of 380~nm which is in good agreement with the geometric
dimensions of our structure. There is also a weaker mode present
at 10~T$^{-1}$ for electrons cycling closer around the inner
diameter of the ring ($D \approx 230$~nm). The two higher
frequency components at 45~T$^{-1}$ and 63~T$^{-1}$ that appear in
Fig.~\ref{cb-ab}(b) are due to electrons that travel twice or
three times around the ring before tunnelling into the drain
contact.

The design of this device with the attached in-plane gates allows
to measure the electrical Aharonov-Bohm~\cite{baumgartner98}
effect as well (data not presented here). It is also possible to
tune the tunnelling barriers into an intermediate coupling regime
to study the Kondo-effect~\cite{goldhaber98} in such quantum
rings.

In conclusion we have fabricated a sub-micron quantum ring
structure with direct nanolithography using an atomic force
microscope. In the open transport regime we observed Aharonov-Bohm
oscillations with amplitudes of more than 50\%. The device was
even tuned into the Coulomb-blockade regime now showing
Aharonov-Bohm like oscillations. These rings are ideally suitable
for detailed studies of phase coherent and interference effects in
ballistic systems from the strong to the weak coupling regime.

We acknowledge financial support from the BMBF. We thank F. Hohls
and P. Hullmann for helpful discussions.

\end{document}